\documentclass[12pt,preprint]{emulateapj}
\usepackage{amsmath, natbib, placeins,color}
\begin{document}

\newcommand\T{\rule{0pt}{2.6ex}}       
\newcommand\B{\rule[-1.2ex]{0pt}{0pt}} 


\title{Is the Ultra-High Energy Cosmic-Ray Excess Observed by the Telescope Array Correlated with IceCube Neutrinos?}
\author{Ke Fang$^{1,2}$, Toshihiro Fujii$^1$, Tim Linden$^{1}$, Angela V. Olinto$^{1,2}$}
\affil{$^1$ The Kavli Institute for Cosmological Physics, The University of Chicago, Chicago, IL 60637, USA}
\affil{$^2$ Department of Astronomy \& Astrophysics, The University of Chicago, Chicago, IL 60637, USA.}
\shortauthors{}
\keywords{(ISM:) cosmic rays --- gamma rays: theory --- gamma rays: observations}

\begin{abstract}
The Telescope Array (TA) has observed a statistically significant excess in cosmic-rays with energies above 57 EeV in a region of approximately 1150 square degrees centered on coordinates (R.A.~=~146.7,~Dec.~=~43.2). We note that the location of this excess correlates with two of the 28 extraterrestrial neutrinos recently observed by IceCube. The overlap between the two IceCube neutrinos and the TA excess is statistically significant at the 2$\sigma$ level. Furthermore, the spectrum and intensity of the IceCube neutrinos is consistent with a single source which would also produce the TA excess.  Finally, we discuss  possible source classes with the correct characteristics to explain the cosmic-ray and neutrino fluxes with a single source.
\end{abstract}

\section{Introduction}
\label{sec:introduction}

\begin{figure*}
                \plotone{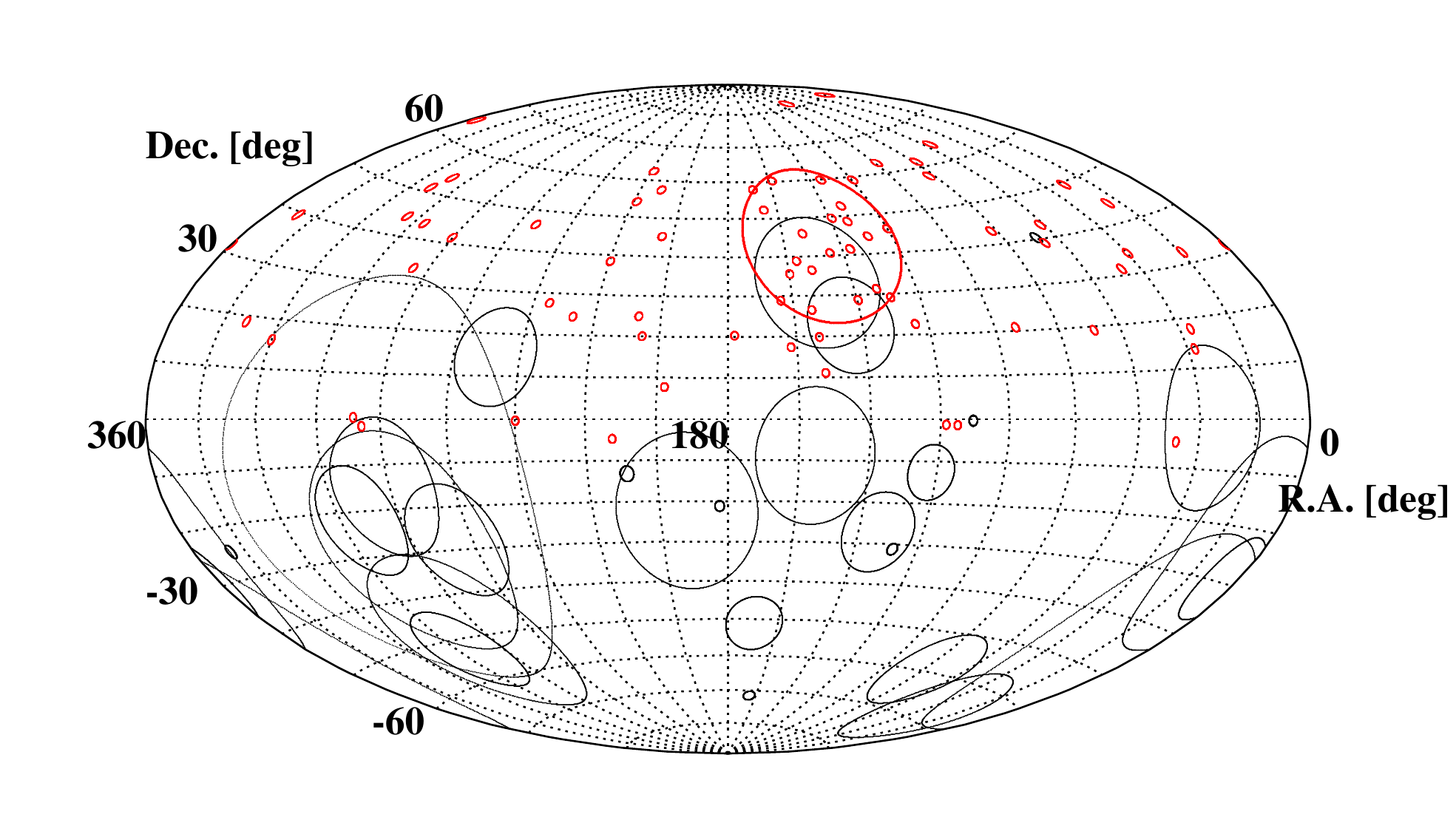}
                	\caption{\label{fig:skymap} Sky map in equatorial coordinates of the 28 IceCube neutrinos (black circles) \citep{2013Sci...342E...1I} and  Telescope Array 5-year events with $E > 57$ EeV and zenith angle $\theta < 55^\circ $  (red small circles) \citep{Abbasi:2014lda}. The TA hotspot is indicated with the $20^\circ$ radius circle of the center of the excess.}
\end{figure*}

The Telescope Array (TA) experiment is the largest hybrid detector in the northern hemisphere advancing the effort to understand the sources of ultra-high energy cosmic rays (UHECRs).   The TA collaboration recently reported an analysis of five years of data measured by the surface detectors in which they find a cosmic ray hotspot with 5.1$\sigma$ significance  within 20$^{\circ}$~radius circle centered at R.A.~=~146.7, ~Dec.~=~43.2 (post trial chance probability $3.4 \sigma$) ~\citep{Abbasi:2014lda}. If this signal is real, it may signal the first detection of an ultra-high energy cosmic-ray (UHECR) point source in the universe. 

Additionally, the IceCube collaboration recently reported the first detections of extraterrestrial neutrinos with energies above 30~TeV~\citep{2013Sci...342E...1I}. These 28 events are distributed across the sky and are currently consistent with an isotropic distribution, although an excess of marginal statistical significance is detected in the southern hemisphere. The effective area of the southern sky is much larger for the IceCube detector, due to the fact that the Earth is opaque to very high energy neutrinos. The effective area of the southern sky is roughly double that of the North at an energy of 100 TeV, and roughly three times as large at an energy of 1~PeV. Additionally there are significant atmospheric backgrounds in the southern hemisphere due to muon showers, which are not present in the north.  Due to these facts, 24 of the 28 IceCube neutrinos have best fit positions in the southern hemisphere, and the effective area of IceCube is relatively small in the region surrounding the TA excess. Interestingly, two  high energy neutrinos observed in the northern hemisphere are found in a similar region as the TA source, out of only four neutrinos observed throughout the northern hemisphere (Fig.~\ref{fig:skymap}). 

High energy neutrinos and UHECRs are known to be tightly connected. 
In addition to EeV cosmogenic neutrinos generated during the UHECR propagation from the source to the earth, TeV - PeV high energy neutrinos can be produced when cosmic rays interact inside the source, if the source has a dense photon background (e.g., AGNs \citep{PhysRevLett.66.2697, Murase:2014foa}, GRBs \citep{Murase:2013ffa}) or hadron background (e.g., newborn pulsars \citep{Fang:2012rx}); and (or) in the source's local environment (e.g., galaxy clusters/groups and star-forming galaxies \citep{Kotera:2009ms, Murase:2013rfa}).

Here we test the correlation between the excess observed by TA and the two spatially coincident events observed by IceCube.  We employ Monte Carlo statistics to demonstrate that the overlap between these two excesses is statistically significant at the 2$\sigma$ level. We then examine the spectrum of the neutrino signal, and find that while huge uncertainties are present, 
the signal is consistent with that observed in UHE.
 Finally, we note that the single source that hosts both the UHECR and neutrino excesses  could be a star-forming galaxy. 

\section{The Spatial Coincidence of IceCube Events}
\label{sec:morphology}

The hotspot observed by TA can be modeled as an ellipse which stretches approximately from an R.A. of 125$^\circ$-170$^\circ$ and a dec. of 20$^\circ$ -- 65$^\circ$\citep{Abbasi:2014lda} (indicated as a red circle in Fig.~\ref{fig:skymap}). This covers an area on the sky of 0.38~sr.  We can compute the probability that two IceCube neutrinos land within this target region. We note that the IceCube collaboration published instrumental effective areas for both northern hemisphere and southern hemisphere searches. They find the effective area in Northern hemisphere searches to fall approximately a factor of 2--3 below that of Southern hemisphere searches, a fact consistent with the observation of 24 neutrinos in the southern hemisphere and only 4 in the Northern hemisphere.

While the effective area of IceCube likely has a complicated spatial dependence, we assume that the effective area is constant within the Northern Hemisphere. While this assumption is naive, the primary contribution to the decreasing effective area in the Northern hemisphere is the amount of matter a neutrino must travel through before reaching the IceCube detector. Thus, the first order correction is that the northern hemisphere effective area will be highest along the equatorial plane, and smallest at high declination. Since the TA excess is centered around dec.~=~43.2$^\circ$, the first order correction to the effective area in this region would be approximately 0, or perhaps slightly negative, making our results conservative. Given a flat effective area, we compute the probability for two IceCube events to have central positions within the TA hotspot to be 0.017, and we can thus reject the coincidence of the IceCube events with the TA excess at 2.1~$\sigma$. Including a 5th neutrino (corresponding to neutrino 5, which is centered at a declination of only -0.4$^\circ$) dilutes this result somewhat to 1.9$\sigma$. We note that additional modeling of the IceCube effective area (as well as the observation of additional neutrinos) will further clarify or rule out this enticing coincidence. 

Furthermore, we note that the statistical correlation between these IceCube neutrinos is also of interest. The IceCube collaboration notes that the morphology of the 28 observed extraterrestrial neutrinos is consistent with isotropy, although a slight excess is observed in the southern hemisphere. However, this finding of isotropy does not necessarily hold for any particular set of neutrinos, once a specific region of interest is picked out a priori. Using the IceCube collaborations published error ellipses for each of its recorded neutrinos, the probability of two northern hemisphere neutrinos being located within their 1$\sigma$ error ellipses is 0.21 (0.8$\sigma$), while the probability of two neutrinos being located closer than the observed offset of 13$^\circ$ is 0.14 (1.1$\sigma$). Though neither of these measures are statistically significant, the fact that the two neutrinos also overlap the TA hotspot is statistically interesting. Further data from the IceCube experiment may be able to soon conclusively determine whether there is an anisotropy in the IceCube neutrino flux which is consistent with the TA excess.

\section{The IceCube Neutrino Spectrum}
\label{sec:spectrum}

\begin{figure*}
                \plottwo{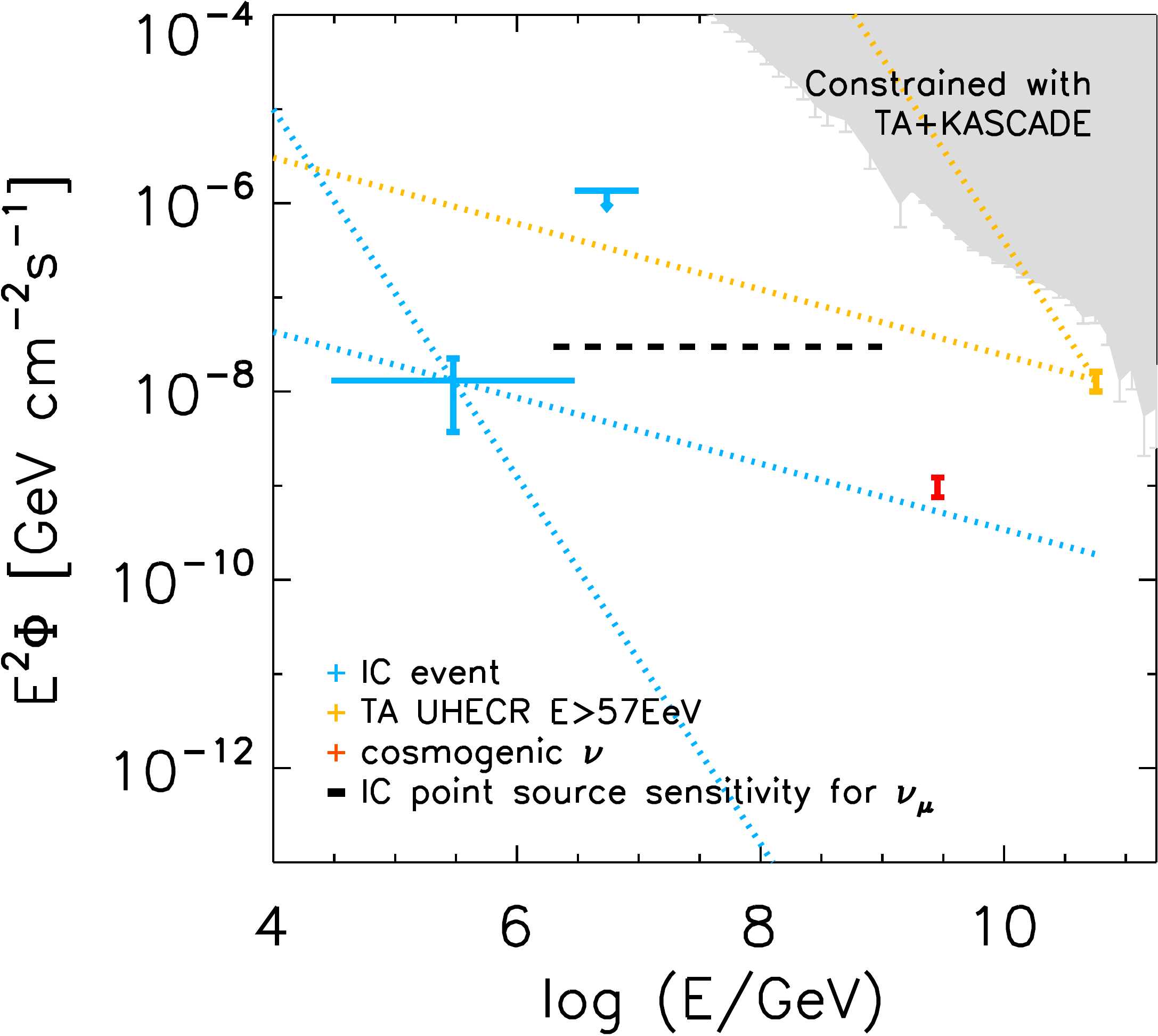}{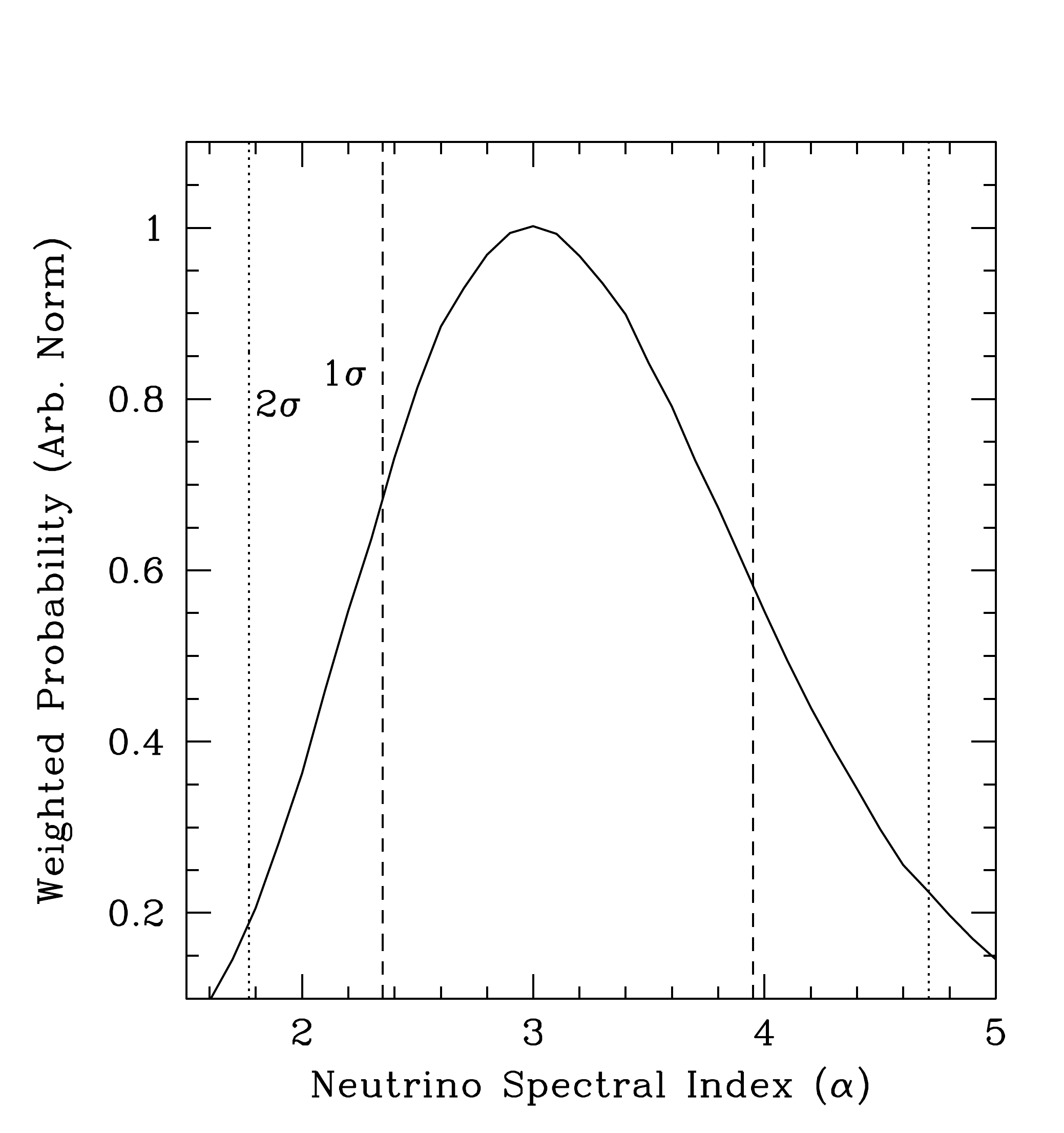}
                	\caption{ \label{fig:1}(Left) The spectrum of a neutrino point source producing events within 20$^\circ$ of the center of the TA excess (blue crosses) compared to the predicted neutrino flux (blue dash lines) given the UHECR intensity observed by TA (orange dash lines). We bin the neutrino spectrum in a single energy bin stretching from 30~TeV - 3~PeV, which spans the entire range of the IceCube detected neutrinos, and place an upper limit on the energy bin spanning 3--20~PeV.  The red data point indicates the maximum GZK neutrino flux the UHECRs could produce. (Right) The weighted probability that a neutrino source with a given spectral index provides only two neutrinos at 63 TeV and 210 TeV, given the energy dependence of the IceCube effective area, along with 1$\sigma$ and 2$\sigma$ error bars for the best fit spectral index. }
\end{figure*}

In addition to checking the spatial consistency of the IceCube and TA data, it is worth checking whether the spectrum and intensity of the neutrino events are consistent with the TA excess. While constraints on the spectrum are extremely loose due to small-number statistics, it is worth asking whether the intensities of each excess are consistent, given theoretical models for the spectra of each messenger. In order to calculate the neutrino intensity, we partition the neutrino events into two energy bins: one which stretches from 30~TeV to 3~PeV, and contains 2 events against a northern hemisphere astrophysical background of approximately 0 events and a  diffuse background of 0.22 events, which is calculated based on the probability that an event from an isotropic direction in the northern sky happens to land inside the TA excess region. Additionally, we list a second energy bin from 3~PeV to 10~PeV, which contains no events and no background. Any flux yielding more than 2.3 events in the high energy bin is thus excluded at more than $90\%$ confidence \citep{Anchordoqui:2013qsi}. 

Adopting the IceCube Northern Hemisphere effective area and 662 days live time of detection, we find the neutrino flux in the hotspot region to be $(1.3\pm0.9)\times10^{-8}\,\rm GeV\,cm^{-2}\,s^{-1}$, as indicated by the blue cross in Fig.~\ref{fig:1} (left). The null bin poses a loose limit because IceCube has a relatively small effective area in Northern Hemisphere above 1~PeV.  This upper limit can be enhanced with the IceCube point source sensitivity at the hotspot declinations, for muon neutrinos with energy from 1~PeV to 1~EeV in $E^{-2}$ spectrum, presented as the black dash line in Fig.~\ref{fig:1} (left) \citep{Aartsen:2013uuv}. We note that these bins were picked in order to encapsulate the entire extraterrestrial IceCube neutrino flux into a single energy bin, and that it is difficult to calculate any spectrum from this process as the choices in binning parameters greatly affects the calculated best fit spectrum. 

Due to larger statistics, the flux of the TA excess can be determined with greater statistical significance. The TA hotspot contains $19\pm4.49$ out of the 72 TA highest energy cosmic ray events, corresponding to a flux above 57~EeV of:  

\begin{equation}
\begin{array}{l l}
E^2J_{H} & = E^2J(E)_{4\pi}\left (\frac{N_{H}}{N_{4\pi}}\right )\left ( \frac{4\pi}{\Omega_{20^\circ}}\right ) \left ( \frac{A_{4\pi}}{A_{H}}\right ) \\
& = (4.4\pm1.0)\times10^{-8}\,\left (\frac{E}{10^{19.5}\,\rm eV}\right )^{-2.6} \,\rm\frac{GeV}{cm^{2}\,s\,sr}
\end{array}
\end{equation}
where $E^3J_{4\pi}=7.9 \times 10^{-9}\,\rm GeV\,cm^{-2}\,s^{-1}\,sr^{-1}$ is the all-sky UHECR flux at $10^{19.5}\,\rm eV$\citep{2013arXiv1301.1703K}, 
 $A_{\rm H}/A_{4\pi}=1/2$ \citep{2013arXiv1310.0647T} and $N_{\rm H}/N_{4\pi}$ are the ratios of effective areas and event numbers between the hotspot region and all sky.

Over a distance $D$ from source to our Galaxy, UHECRs are expected to experience a deflection angle $\delta\theta_{\rm EG}\approx 3-5^\circ\, Z\,(D/100{\rm \,Mpc})^{1/2}\,(E/100{\,\rm EeV})^{-1}$ in a  turbulent extragalactic magnetic field (EGMF) with field strength $B_{\rm EG}\sim 2\,\rm nG$ and coherence length $\lambda_{\rm EG}\sim 300\,\rm kpc$ \citep{MiraldaEscude:1996kf}. The Galactic magnetic field (GMF) adds another $\delta\theta_G\approx1-5^\circ\,Z$ depending on the arrival direction of the events \citep{Giacinti:2010dk, Farrar:2012gm}.  This is somewhat smaller than the observed extent of the TA hotspot region. However, crossing  the intergalactic magnetic field can further enhance the deflection between the arrival direction of super-GZK
protons and the sky position of their actual sources  to be $\sim 15^\circ$ \citep{Ryu:2009pf}. Additionally, if the UHECR flux is dominated by higher-Z particles, as
suggested by data from the Pierre Auger Observatory \citep{Abraham:2010yv, Aab:2013ika},  then the deflection angle would be enhanced by a factor of Z. Thus, the large diffuse region of the TA hotspot is entirely consistent with emission from a single bright point source.
Interestingly, we note that the existence of a single bright UHECR source in the northern hemisphere could help elevate the tension between the chemical composition measurements (Xmax and RMS-Xmax) of TA and Auger \citep{Array:2013dra}. Specifically, while TA observed a UHECR flux compatible with 100\% protons, observations by Auger indicate the presence of heavy nuclei in the UHECR flux. 
If this bright UHE source is a proton accelerator, it would strongly contribute to the TA Xmax measurements, compared to the chemical composition of the isotropic background. Alternatively, if the source emits both light and heavy nuclei and if the EGMF and GMF are intense, the protons from this source will be less deflected by the  magnetic fields, while the heavy nuclei will be greatly deflected. This would produce an overabundance in light nuclei in the TA region of interest (ROI), and an overabundance of heavy nuclei in the Auger ROI. Future gamma-ray observations of the hotspot region can help constrain the strength of the magnetic fields in order to constrain the scenarios.

If the UHECR source powering the TA hotspot is at large distance, cosmic rays above the GZK energy would lose energy via photo-pion interaction on cosmic microwave background in their extragalactic propagation, and produce cosmogenic neutrino flux up to $E_\nu^2\Phi_\nu=3/8\,E^2J(E)_{\rm H}\,f_\pi=(1.0\pm0.2)\times10^{-9}\,\rm GeV\,cm^{-2}\,s^{-1}$, assuming  on average each pion takes $20\%$ proton energy and a maximum pion production rate $f_\pi=1$.  As the red mark in Fig.~\ref{fig:1} (right) shows, this flux level is well below the high energy IceCube upper limits, suggesting that  it is consistent with both UHECRs and neutrinos  emitted by the same source.

In Figure~\ref{fig:1} (right), on the other hand, we attempt to constrain the spectrum of the neutrino source in a bin-independent way. It is worth noting that the neutrino spectrum can be constrained not only from the two observed events at 63 TeV and 210 TeV, but also by the lack of detections at any other energy. We adopt a flat prior on the spectral index and then produce a Monte Carlo population of neutrinos for each spectral index. Weighting the spectrum by the energy dependent effective area of the IceCube experiment in the northern hemisphere~\citep{2013Sci...342E...1I}, we calculate the relative likelihood that emission spectra with varying spectral indices would produce the observed data. Given the existence of only two events, the constraints on the spectral index are extremely loose, with value of 3.0, and a 1$\sigma$ (2$\sigma$) confidence interval of 2.35~--~3.95 (1.77~--~4.71). 

Using this wide range of spectral indices, we can attempt to determine whether the neutrinos observed at TeV energies may be correlated with the UHECR observed by TA. If we assume that all cosmic rays are accelerated with the same mechanism with a single power-law, to a zero-order approximation, the neutrinos would be expected to follow the same spectral index till the cosmic ray energy hits the pion production threshold. 

In Figure~\ref{fig:1} (left) we show blue dashed lines indicating the possible neutrino spectral indices for fits at our 1$\sigma$ and 2$\sigma$ confidence intervals, along-side orange-dashed line indicating the corresponding cosmic-ray spectrum. We note that while our neutrino data is centered around a spectral index $\alpha$~=~3.0, cosmic-ray all-sky observations by TA \citep{2013arXiv1301.1703K} and KASCADE \citep{Apel:2013ura} disfavor indices softer than 3.0. A spectral index between 2.0 and 3.0 is highly consistent with neutrino and UHECR observations and produces a coherent picture for all data. 



\section{ Possible Single Source Scenarios?}
\label{sec:mkr421}

\begin{figure}
                \plotone{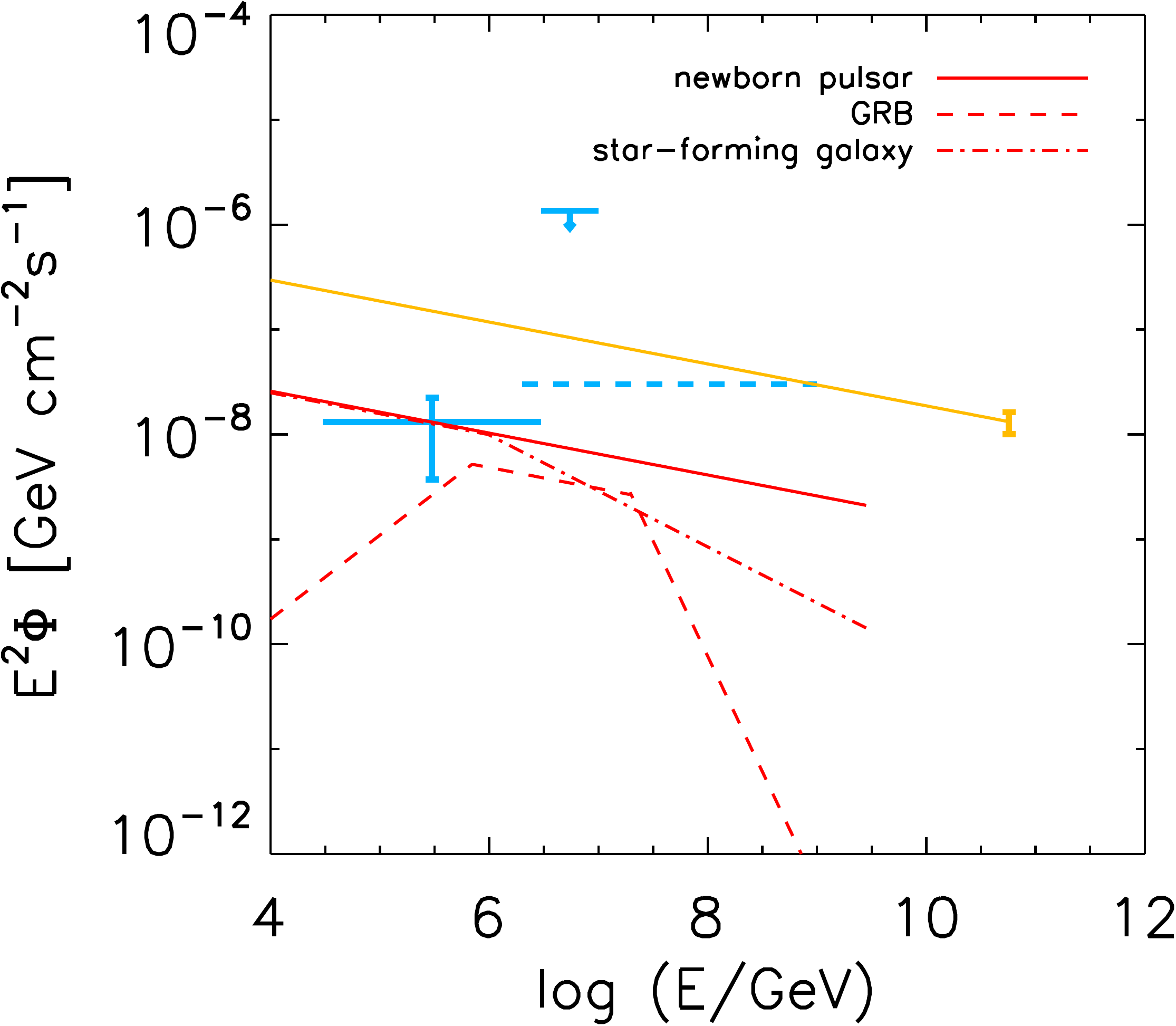}
                	\caption{\label{fig:3} 
{
{The fit of several astrophysical source classes (newborn pulsar, GRB, star-forming galaxy) to the spectrum and intensity of the IceCube neutrino flux (shown in blue), assuming a cosmic-ray injection spectrum falling as $E^{-2.2}$ and normalized at $57\,\rm EeV$ to the UHECR intensity of the TA excess (shown in orange). Each source class is found to be consistent with both the UHECR and neutrino excesses assuming standard model parameters, which are described in detail in the text.}
}
}
\end{figure}

If the IceCube neutrinos and TA hotspot are indeed related, then it is worth considering possible sources of UHECRs which could produce both signals. While it is possible that these sources are produced by an over-density in extragalactic UHECR production in this direction, the high fractional intensity of the TA excess coupled with the relative smoothness of the Fermi-LAT extragalactic $\gamma$-ray sky disfavor this scenario and instead indicate emission from either a single bright UHECR source, or possibly a handful of such sources~\citep{2010PhRvL.104j1101A, 2012PhRvD..85h3007A}.

Any potential source of both the TA and IceCube excesses should be able to power both UHECRs and TeV-PeV neutrinos. In addition, it should not overproduce PeV-EeV neutrinos compared to the point-source limits produced by IceCube (blue dash line in Fig~\ref{fig:3}). 
Below we investigate a few possible scenarios in detail and summarize them in Table~\ref{table:1}.


Gamma-ray Bursts (GRBs) have long been suggested as potential UHECR sources (see \cite{0034-4885-69-8-R01} for review), and high energy neutrino emission peaked around PeV can happen when cosmic rays interact with fireball photons \citep{Waxman:1997ti, Murase:2005hy,2013JCAP...06..030C,  2013ApJ...766...73L, Murase:2013ffa}. The non-thermal photon spectra of GRBs can be
described by a broken power law, $n(\epsilon) \propto \epsilon^{-\alpha}$ below the break energy $\epsilon_b$  and $n(\epsilon) \propto \epsilon^{-\beta}$ above, with $\alpha\sim 1$ and $\beta\sim 2$. The photopion production efficiency is 
\begin{eqnarray}
f_{p\gamma}(\epsilon)  &=& 0.2\,\frac{L_{\gamma,51}}{\epsilon_{b,\rm MeV}\,\Gamma_{2.5}^4\,\delta t_{\rm ms}}  \\
&\times&
\begin{cases}  (\epsilon/\epsilon_{b})^{\beta-1}  & \epsilon \le \epsilon_b
\\
 (\epsilon/\epsilon_{b})^{\alpha-1}& \epsilon > \epsilon_b \end{cases} 
\end{eqnarray}
for a GRB with luminosity $L_{\gamma,51} = L_{\gamma} / 10^{51} \rm erg\,s^{-1}$, bulk Lorentz factor $\Gamma_{2.5} =\Gamma/10^{2.5} $  and variability time $\delta t_{\rm ms} = \delta t/1\,\rm ms$ (corresponding to $R = 2\Gamma^2\,c\,\delta\,t \approx 10^{13}\,\rm cm$).  A photon break energy in observer frame at $\epsilon_{b, \rm MeV} = \epsilon_b/1\,\rm MeV$ corresponds to a break in neutrino spectrum  at $E_\nu^b = 7\times10^{14}\, \Gamma_{2.5}^2\,\epsilon_{b,\rm MeV}^{-1}\,\rm eV$.

 The neutrino spectrum of a GRB can be approximated by
\begin{eqnarray}
E_\nu\,\Phi_\nu  &=& \frac{3}{8}\,f_{p\gamma}\,E_{\rm CR}\,J_{\rm CR}  \\
&\propto&
\begin{cases} (E_\nu/E_\nu^b)^{\beta+1 -s} & E_\nu \le E_\nu^b
\\
(E_\nu/E_\nu^b)^{\alpha+1 -s} & E_\nu^b < E_\nu < E_\nu^c
\\
(E_\nu/E_\nu^c)^{\alpha-1 -s} &E_\nu > E_\nu^c
\end{cases}
\end{eqnarray}
Here  $E_\nu^c = 2\times10^{16}\, L_{\gamma,51}^{-1/2}\,\Gamma_{2.5}^4\,\delta t_{\rm ms}\,\rm eV$. This additional break in the neutrino spectrum is caused by the synchrotron cooling of charged pions in the magnetic field of the dissipative fireball, and the ratio of synchrotron cooling time to pion decay time gives a suppression factor $\sim E_\nu^{-2}$. Assuming an intrinsic cosmic-ray spectrum $E^{-2.2}$ normalized by the TA excess, a GRB with typical parameters is capable to produce the  neutrino excess   (as shown by  the red dashed line in Fig~\ref{fig:3}). 


Another possible channel of producing TeV-PeV neutrinos is via hadronuclear interaction (pp or Np) in baryon-rich sources, including newborn pulsars \citep{Fang:2012rx, Fang:2013vla} and magnetars \citep{Arons:2002yj, 2009PhRvD..79j3001M}, galaxy clusters \citep{Inoue:2005vz, Kotera:2009ms, Murase:2013rfa} and  star-forming galaxies \citep{Loeb:2006tw, Katz:2013ooa, Murase:2013rfa, Tamborra:2014xia, Anchordoqui:2014yva}. The neutrino spectrum from hadronuclear interaction inherits the cosmic-ray spectrum, $E_\nu^2 \Phi_\nu = 3/8 \, f_{\rm pp} \, E_p^2\,J_p$. In order to connect the flux of the neutrinos and the UHECRs of the excess, a pion production efficiency $f_{\rm pp}\sim 0.2$ is assumed for an intrinsic UHECR spectrum falling as $E^{-2.2}$. Specifically,  $f_{\rm pp} =  n_p\,\sigma_{\rm pp}\,\kappa_{\rm pp}\,c\,t_{\rm int}$, where $n_p$ is the target baryon density, $t_{\rm int}$ is the duration of the interactions and $\sigma_{\rm pp} \sim 10^{-25}\,\rm cm^2$ and $\kappa_{\rm pp}\sim 0.5$ are the proton-proton interaction cross section and elasticity. 

In pulsars and magnetars, the baryon target is the surrounding supernova ejecta.  For an ejecta mass $M_1=M/10\,M_\odot$, the expansion speed is $\beta_{-1.5}\,c=10^9\,\rm cm\,s^{-1}$, $f_{\rm pp} = 0.2\,M_1\,\beta_{-1.5}^{-2}\,t_{\rm yr}^{-2}$ for a duration $t_{\rm yr} = t/1\,\rm year$.   The neutrino production from a newborn pulsar with surface magnetic field $10^{12}\,\rm G$ and initial spin period $1\,\rm ms$ is indicated as the red solid line in Fig~\ref{fig:3}. Notice that the typical time scale for a magnetar is $t\approx 300\,\rm s$, so while UHECRs can be accelerated, they are unlikely to escape from the source \citep{Fang:2012rx}. 

Alternatively in galaxy clusters, cosmic rays can be accelerated by the accretion shocks formed around large scale structures, and then subsequently interact with nucleons of the intergalactic medium. A typical intergalactic gas density of $n\sim 10^{-4}\,\rm cm^{-3}$ \citep{Voit:2004ah} yields a value of $f_{\rm pp} $ ranging from  $10^{-4} - 10^{-2}$\citep{Kotera:2009ms}. Hence the neutrino flux from a single galaxy cluster providing the TA UHECR flux is far below the IceCube detection threshold.

In star-forming galaxies, cosmic rays can be accelerated in the collisionless shocks produced by supernova explosions, and/or by the high energy accelerators contained in the galaxy. The starburst interstellar medium then serves as the baryon target with typical column density $\Sigma_{g,-1}=\Sigma_{\rm g} / 0.1 \,\rm g\,cm^{-2}$, corresponding to a pion production time $t_{\rm pp} = 3\,\Sigma_{g,-1}^{-1}\,H_{\,\rm kpc}\,\rm Myr$. The time cosmic rays stay in the galaxy, $t_{\rm esc}$, depends on energy.
At low energy cosmic rays are  confined in the  starburst-driven wind and transport via advection, so $t_{\rm esc}\approx 10\,H_{\rm kpc}\,v_{g,7}^{-1}\,\rm Myr$ for a galaxy with scale height of $\sim 1\,\rm kpc$ and wind speed $v_{g,7}=v_g/100\,\rm km\,s^{-1}$.  These particles undergo strong  interactions as $f_{\rm pp}=t_{\rm esc}/t_{\rm pp}\sim 1$.
At higher energy  diffusion  dominates, then $t_{\rm esc}\approx H^2/4D = 1.6\,D_{0,26}^{-1}\,H_{\rm kpc}^2\,(E/100\,\rm PeV)^{-1/3}\,\rm Myr$, assuming Kolmogorov turbulence and diffusion coefficient $D_0\approx 10^{26}\,\rm cm^2\,s^{-1}$ at GeV  \citep{Murase:2013rfa}. A spectral break at $E_\nu^b\approx 1\,\Sigma_{g,-1}^{3}\,H_{\rm kpc}^3\,D_{0,26}^{-3} \,\rm PeV$ is expected as  cosmic-rays with higher energy are less confined for interactions. 
This case is shown as red dash-dotted line in Fig~\ref{fig:3}.


Another important class of high energy neutrino sources is active galactic nuclei (AGN) \citep{PhysRevLett.66.2697, Winter:2013cla, Murase:2014foa, Dermer:2014vaa}. In BL Lac objects, non-thermal photon emission from the inner jet is the dominant interaction target for UHECRs. The photomeson production rate is  $f_{\rm p\gamma} = 10^{-4}\, L_{\gamma,45} \, t_5^{-1} \, \Gamma_1^{-4}\, (\epsilon_b / 100\,\rm eV)$ at time $t_5 = t/ 10^5\,\rm s$ in a jet with Lorentz factor $\Gamma_1 = \Gamma/10$ and gamma-ray luminosity $L_{\gamma,45}=L_\gamma/10^{45}\,\rm erg\,s^{-1}$ \citep{Murase:2014foa}. The TA excess is located in the direction of Markarian 421, which is among the best candidates for a blazar producing UHECR. This is due to the fact that Mrk 421 is both among the brightest $\gamma$-ray blazars \citep{2011ApJ...736..131A} and is relatively close ($\sim$134 Mpc~\citep{2012AJ....143...23G}), allowing UHECRs to reach Earth at energies above the GZK cutoff. 

However, given the low pion production efficiency of Mrk 421, it is difficult for this source to produce the bright neutrino flux observed by IceCube \citep{Dimitrakoudis:2013tpa}. One caveat to this analysis is the addition of thermal gas in the source region, which would make it more opaque to UHECRs. However, this is difficult to reconcile with X-Ray/TeV observations that suggest the ratio of thermal and synchrotron photons to be $\sim0.1$ \citep{Fossati:2007sj}).
One possible solution is an existence of high-energy hardening in the UHECR spectrum, significantly increasing the PeV cosmic-rays responsible for the IceCube neutrino flux. 

In contrast to AGN such as Mrk 421, flat spectrum radio quasars (FSRQs) provide a photopion production efficiency of $ 1 - 10\%$ due to interactions with photons of the broad-line region and the accretion disk,  but the cosmic-ray spectrum must be  softened above $10^{16}\,\rm eV$ in order to avoid overproduction of EeV neutrinos. Therefore FSRQs cannot be significant UHECR sources \citep{Dermer:2014vaa}.

\begin{table}[ht]
\caption{SUMMARY OF SOURCE POSSIBILITY \footnote{estimated at typical parameters} } \label{table:1}
\centering
\begin{tabular}{c c c c}
\hline\hline
Source & Can host & Can produce &Reasonable  \T \\ 
Type &  TA excess? & the two $\nu$s? &  Associations?  \B \\ 

\hline
GRB &Y & Y & Transient \T \\
Star-forming galaxy & Y & Y & Y \\
Fast-spinning Pulsar & Y & Y  & Transient\\
Magnetar & N & Y  & Transient\\
Galaxy Cluster & Y & N & Y\\
BL Lac Object & Y & N & Y\\ 
FSRQ & N & Y \B & N\\  
\hline
\end{tabular}
\end{table}

We summarize the above discussions in the first three columns of Table~\ref{table:1}.
In the fourth column we indicate the availability of such sources within $20^\circ$ of the Hotspot center. Notice that GRBs and pulsars are transient sources, and their neutrino emission time ($\delta t_\nu^{\rm GRB} \sim \rm sec$, $\delta t_\nu^{\rm pulsar}\sim \rm yr$) is much shorter than the dispersion time in UHECR arrivals, $\delta t_{\rm UHECR} \approx 4\times10^4\,(l/100\,\rm Mpc)\,(\delta\alpha/1^\circ)^2\,\rm yr$ for every 100 Mpc propagation and $1^\circ$ deflection in the IGMF \citep{Kotera:2007ca}. Therefore time correlation between neutrino and UHECR arrivals is not expected for these types of sources, unless that  sources are born periodically in certain star-rich region. On the other hand, for the steady types, we investigate the relevant sources  in the available high-energy catalogues and surveys \citep{Gao:2003qp,2011A&A...534A.109P,Ackermann:2012vca,2013arXiv1306.6772T,::2013ufa}. We 
 list a selection of results in Table~\ref{table:2}. Future measurements will narrow down the candidates, with high energy neutrino observation limiting or confirming the source direction(s), and more statistics of UHECRs  helping to constrain the source distance(s) and other properties.

\begin{table*}[t]
\caption{List of high energy sources around the TA hotspot and IceCube neutrino 9 \& 26 } \label{table:2}
\centering
\begin{tabular}{c c c c c c c c c}
\hline\hline
Source & Catalogues & RA & Dec & offset to & offset to & offset to & Distance & Source \T  \\ 
Name &  	/ Surveys & (J2000)	 & (J2000) 	 & HS center  & $\nu$9  & $\nu 26$ & (Mpc) & Type \B   \\ 
\hline
MKN 421\footnotemark[1] & TeVCat\footnotemark[5] /1FHL  & 11 04 19.0 & +38 11 41 &   15.5&  12.8&  24.8& 134 &
 BL Lac \T \\ 
1ES 1011+496 \footnotemark[1]& TeVCat/1FHL & 10 15 04.0 & +49 26 01 &      7.9&  15.9&  28.0& 1036 &
 BL Lac \\

Arp 55\footnotemark[2] & Fermi/HCN Survey & 09 15 55.2 &  +44 19 55 
&     5.7&  14.4&  21.9& 162.7 &
star-forming galaxy \\

NGC 2903 \footnotemark[2]&Fermi/HCN Survey &  09 32 09.7 & +21 30 02 
 &    21.9&  14.1&   1.2& 6.2 &
star-forming galaxy \\

UGC 05101 \footnotemark[2]&Fermi/HCN Survey & 09 35 51.6 &  +61 21 12
&    18.2&  28.2&  38.7& 160.2 & 
star-forming galaxy \\

M82 \footnotemark[2]&Fermi/HCN Survey/TeVCat &  09 55 51.6  &+69 40 45
 &  26.5&  36.1&  47.1& 3.4&
star-forming galaxy\\

NGC 3079 \footnotemark[2] &Fermi/HCN Survey& 10 01 57.8  &+55 40 49
 &  12.7&  22.1&  33.4& 16.2 &
star-forming galaxy \\

IRAS 10566 \footnotemark[2]&Fermi/HCN Survey& 10 59 18.2 & +24 32 34
&    23.9&  14.9&  19.7& 173.3 &
star-forming galaxy \\

Arp 148 \footnotemark[2] &Fermi/HCN Survey& 11 03 53.7 & +40 50 59
&    14.5&  13.7&  26.3& 143.3 &
star-forming galaxy\\

NGC 3556 \footnotemark[2]&Fermi/HCN Survey& 11 11 31.8 & +55 40 15
 &    18.4&  24.9&  37.6& 10.6 &
star-forming galaxy \\

NGC 3627 \footnotemark[2]& Fermi/HCN Survey& 11 20 14.4 & +12 59 42
 &   36.3&  26.8&  27.1& 7.6 &
star-forming galaxy \\

NGC 3628\footnotemark[2] & Fermi/HCN Survey& 11 20 16.3  &+13 35 22
 &      35.8&  26.3&  26.9& 7.6&
star-forming galaxy \\

NGC 3893 &Fermi/HCN Survey&  11 48 39.1 & +48 42 40
&   21.7&  24.4&  37.2& 13.9 &
star-forming galaxy  \\

A1367 (Leo)\footnotemark[3] & HIFLUGCS & 11 44 28.8 &  +19 50 24 &   33.9&  26.0&  30.6& 87.5 & Massive galaxy cluster \\
A1035 \footnotemark[4] & MCXC& 10 32 14.8 & +40 15 53 &    9.0&   8.6&  21.5& 317.7 & galaxy cluster \B \\

\hline
\end{tabular}
\footnotetext[1]{Listed as one of the most probable counterparts of the IceCube high-energy neutrinos in \cite{Padovani:2014bha}  }
\footnotetext[2]{Selected starburst galaxies detected by FERMI LAT gamma-ray telescope\citep{Ackermann:2012vca}, sample based on the HCN survey of \cite{Gao:2003qp}}
\footnotetext[3]{One of the three galaxy clusters with gamma-ray excess at a post-trial significance of  $2.6 \sigma$ \citep{::2013ufa}}
\footnotetext[4]{An example of nearby X-ray clusters selected from MCXC catalog \citep{2011A&A...534A.109P}}
\footnotetext[5]{http://tevcat.uchicago.edu}
\end{table*}

\section{Conclusions}
\label{sec:conclusions}

The recent observation of a UHECR hotspot by the TA observatory has potentially opened the doors to a new era of UHECR point source detections. Interestingly, the TA hot spot also correlates with a marginally statistically significant overabundance in the IceCube neutrino flux. We find that the morphology and spectrum of the TA and IceCube observations are consistent with each other. Moreover, among possible  scenarios, a single source with the type of star-forming galaxy can successfully host both excess at the same time. 

Again, we caution that the correlations presented in this work are highly tenuous, given the extremely small statistics in both the IceCube and TA datasets. However, this fact conversely makes the possible detection of a single point source appealing, since upcoming data releases by IceCube and TA should  confirm or rule out this detection. 

{\it
Shortly after we submitted this work, the IceCube Observatory  updated the high energy neutrino list obtained from unbinding the 2012 - 2013 data \citep{Aartsen:2014gkd}. Three additional events were found in the north hemisphere (not including event 37 which is a track event with very low deposition energy and is highly possible to be a background event). With the 3-year IceCube data the significance of event 9 and 26 overlapping with the TA excess drops to $1.6\,\sigma$, but the conclusions of this paper remain unchanged. 
}

\acknowledgments
We would like to thank Markus Ahlers, Jacob Feintzeig, Jeff Grube, Dan Hooper, Albrecht Karle, Kohta Murase and Nathan Whitehorn for helpful conversations. We also acknowledge comments by the anonymous
referees, which lead to significant improvements in this manuscript.
KF and AVO acknowledge financial support from NASA 11-APRA-0066 and NSF grant PHY-1068696 at the University of Chicago. 
TF is supported by Grand-in-Aid for the Japan Society for the Promotion of Science Fellowship for Research Abroad.
TL is supported by the National Aeronautics and Space Administration through Einstein Postdoctoral Fellowship Award Number PF3-140110. 
KF, TL and AVO acknowledge support from the Kavli Institute for Cosmological Physics through grant NSF PHY-1125897 and an endowment from the Kavli Foundation.
 \\ 

\bibliography{ta_icecube}

\end{document}